\documentstyle[aps,preprint,epsfig]{revtex}

\def\ee{$e^+e^-~$}

\def\Qq{$Q \overline{q}~$}
\def\qQ{$q \overline{Q}~$}
\def\QQ{$Q \overline{Q}~$}

\def\qq{$q\overline{q}~$}
\def\cc{$c\overline{c}~$}
\def\bb{$b\overline{b}~$}

\def\QqqQ{${Q\overline{q}q\overline{Q}}~$}

\begin{document}

\preprint{UCLA/94/TEP/47;
          hep-ph/9412292}
\title{Threshold production of charmed and B mesons in \ee annihilation
\footnote{
 Presented at the International Conference on
QUARK CONFINEMENT AND THE HADRON SPECTRUM,Villa Olmo -- Como, Italy,
 20-24 June 1994. To be published in the Proceedings of the Conference.}}
\author{Nina Byers}
\address{Physics Department, UCLA, Los Angeles, CA 90024}
\date{December 9, 1994}
\maketitle
\begin{abstract}
After an historical introduction reviewing the successes and failures
for heavy quarkonium spectroscopy
of the nonrelativistic quark model including $(v/c)^2$ corrections,
the discussion is widened to include light quark pair creation; i.e.,
dynamical quarks.  We find
 that the simple extension of the QCD inspired
potential model which
 includes dynamical quarks  first proposed by Eichten et al. \cite{eichetal}
is remarkably successful in accounting both for the observed narrow
state heavy quarkonium spectroscopy {\it{and}} experimental data on
 threshold production of charmed and
B mesons in \ee annihilation.  We have studied various different
models of light quark pair creation.  In addition to the original Cornell
model, we find that a variant of it can also account for the data.
The cross section data are mainly inclusive cross section measurements.
The two models give distinguishably different predictions for the
energy dependence of
production cross sections for individual channels, and  measurement of
the energy dependence of production cross sections for individual
channels may distinguish between these models.

\end{abstract}

\section{Introduction}

In this talk I would like to discuss three sorts of QCD inspired
phenomenological models which describe \QQ physics at low energies.\cite{qq}
The simplest is the nonrelativistic naive quark model (NR NQM).  By naive
quark model (NQM), I refer to treatments in which the problem is treated
strictly as a two body problem.  As I remind you below,
the NR NQM initially had many successes.  However, there were also
some failures which were put right when $(v/c)^2$ corrections were
taken into account.  I will discuss separately two types of $(v/c)^2$
corrections; (i) those applied to the NQM and (ii) extension of the NQM
to take into account light quark pair creation.  This is one of the two
main points of my talk, namely that light quark pair creation effects
are $(v/c)^2$ corrections to the NQM and are of the same order as
effects due to the spin dependence of the \QQ force and kinematic
$(v/c)^2$ corrections.  All these effects are important in obtaining
good agreement of quark model predictions with experimental data.
There are two types of light quark pair creation effects to be discussed
here; namely, (1) those that occur in heavy quarkonium
states below flavor threshold
(vacuum polarization effects) and (2) those responsible for
production of charmed and B mesons at low energies. The effects of  virtual
light quark pairs on the
narrow \QQ states below flavor threshold
can be included through modification of the two-body \QQ
potential. If one confines one's attention only to these states, then the
phenomenological \QQ potential obtained by fits to data includes both the
two-body potential and the effect on the spectrum of virtual \qq pairs.
Such a treatment, however, does not allow for relating the effects of
virtual \qq pair creation with actual \qq pair creation as seen in the
productions of charmed and B mesons.  It is also somewhat inadequate in
that it is difficult to obtain the configuration mixing necessary to
account for certain spectroscopic results
without explicit inclusion of four-body \QqqQ virtual states.

In section II is a brief summary of the successes,
and failures, of the NR NQM; and,  in section III,
the improvement when $(v/c)^2$ corrections and the spin-dependence of the \QQ
interaction are taken into account.
Section IV presents some QCD inspired  potential
models of light quark pair
creation. Section V outlines the method used to include these
dynamical quarks in calculating both the spectroscopy of the
narrow \QQ states below flavor threshold and the production cross sections
for charmed and B mesons above flavor threshold. Of the
various possible models we studied in detail,
two seem reasonably  successful in accounting for both the observed
spectroscopy of the narrow states and the observed
threshold production cross sections in \ee annihilation.
These models are the Cornell model \cite{eichetal}
 and an extension of it studied by Zambetakis \cite{zamb}.
 Section VI compares predictions of these models
with available production cross section data. These are mainly includive
cross section data.  Both models give remarkably good accounts of these data
but neither are precise fits to the data.  This section also presents
exclusive production cross sections; for these the models predict different
energy dependences. Measurement of exclusive production cross
sections may indicate which of these models is the more correct one.

\section{Successes and Failures of the  Nonrelativistic Naive Quark Model
 (NR NQM).}
\label{sec:sf}

This section is something of an historical account which is incomplete and
included here to give background to the assertion that
corrections to the NR NQM  owing to light quark pair
creation, and the coupled channel mixing it induces, are of the same
order as those due to spin-dependent forces and other $(v/c)^2$
corrections.

\subsection{Successes}
\label{subsec:snrnqm}

Here are outlined the early outstanding
successes of the NR NQM. Using a simple QCD inspired potential which
behaves at short distances like a one gluon exchange potential and
is linearly rising at large distances, the model has as parameters
the strength of the Coulombic part and the slope of the linear part of
the potential; in addition, the masses of the c and b quarks.  With
these few parameters, the model accounts quite well for
a large number of data:
\begin{quote}
  Masses of narrow heavy quarkonium states (spin averaged).

  Leptonic decay rate ratios $\Gamma_{ee}(\psi')/
\Gamma_{ee}(\psi)$, etc..

Allowed radiative transition rates (given by $<r>_{fi}$); e.g.,
$\psi' \rightarrow \chi_J,~ \chi_J \rightarrow J/\psi$, etc..

  Allowed M1 radiative transition rates (given by $<1>_{fi}$)~; e.g.,
 $J/\psi \rightarrow \eta_c$~.
\end{quote}

\subsection{Failures}
\label{subsec:fnrnqm}

Among the successes above were some failures.  These
were:
\begin{quote}
  $\Gamma_{ee}(J/\psi)$ too big.

  E1 radiative transition rate for $\psi' \rightarrow \chi_0$ too big
by a factor 2.

Forbidden M1 transitions such as $\psi' \rightarrow \eta_c$ observed.

  $\psi(3770)$ observed in \ee annihilation with mass too close to the
$\psi'$ to be
understandable as a $^33S_1$ \cc state.
\end{quote}
In addition this simple model failed to account for the observed fine
and hyperfine splittings in the spectra.  The items above, and the fine
structure of the mass spectra, are corrected by taking relativisitc
corrections into account.

\section{Relativistic  Corrections.}
\label{sec:vccorr}

There are two types of relativistic corrections: (i) $(v/c)^2$  corrections
to the NQM; i.e., $(v/c)^2$ corrections to the nonrelativistic potential
model in which nevertheless the model remains a two-body treatment of the
problem, and (ii) extension of the model to take into account dynamical
quarks; i.e., light quark pair production. Type (i) relativistic corrections
have been discussed in this Conference explicitly by G. M. Prosperi,
Yu-Bing Dong, L. Fulcher, F. Schoeberl, H. Sazdjian, Kuan-Ta Chao,  and
extensively in the literature; see,e.g., \cite{mccl,bram}  and references
cited therein. Spin-dependent forces are also regarded as
type (i) $(v/c)^2$ corrections
because they arise in this order in the Breit-Fermi reduction
of a particle exchange diagram. Estimates of the spin-dependence of the \QQ
force  have been obtained other ways as well; see, e.g.,
\cite{eich-fein,gromes} and the contributions of
J. M. Ball and F. Zachariasen to this Conference.
The effects that I would classify as type (ii) relativistic corrections
are those that arise owing to virtual light quark pairs.  These are of the
same magnitude as those of type (i) and  arise in  relativistic
quantum field theory in order $(v/c)^2$. Below is a list of
 relativistic corrections
which have been found to correct
the 'failures'  mentioned above. In most cases several effects contribute.

\begin{quote}
  a. spin-orbit \QQ interaction

  b. spin-spin \QQ interaction

  c. tensor forces

  d. spin-independent $(v/c)^2$ corrections to the \QQ Hamiltonian

  e. direct \QQ  $^3D_1$  - photon coupling

  f.  light quark pair creation

  g.  coupled channel mixing.

\end{quote}

In Table I. are indicated which of the above items tend to correct the
so-called failures.

\begin{center}\mbox
{\begin{tabular}{l|l}
\hline\hline
\multicolumn{1}{c}{put right} & \multicolumn{1}{c} {by}\\
\hline
Fine structure splitting & ~ a\\
E1 rate $\psi' \rightarrow \chi_0$ &~ a and g \\
$\Gamma_{ee}(J/\psi)$ & ~ g \\
Forbibben M1 rates  & ~ d and g \\
$\psi(3770)$ explained & ~ c, e, and g \\
$\sigma$(\ee $\rightarrow$ charmed mesons) & ~ f \\
\end{tabular}}
\end{center}

\begin{center} {\bf{Table I.}} \end{center}
I have not included hyperfine splittings in the above discussion. This is
because I believe that remains an unsolved problem.

The bulk of this paper is about light quark pair creation. The foregoing
was included to indicate that inclusion of light quark pair creation
improves the the agreement of NQM  results with experimental data in
two ways. Not only does it extend the model to describe quarkonium decays to
charmed and B mesons, it also improves the agreement with the
narrow state spectroscopic data.

Before addressing explicitly light quark pair creation effects, it is
appropriate here to remind you that relativistic effects
such as spin-orbit couplings,
as well as pair creation, reflect the Lorentz
transformation properties of the interaction. For example,
in a Breit-Fermi Hamiltonian  the spin-orbit
interaction coming from vector and scalar exchange have opposite sign and
differ by a factor 3.
 Correspondingly pairs created from
vector and scalar exchange are produced in $1^{--}$
and $0^{++}$ states, respectively.  At present
it is not clear that the spin-orbit interaction is correctly viewed as
an effect found in a Breit-Fermi reduction of
vector and scalar exchange interactions.
However, when this assumption is made and the result compared
with data the observed fine structure
splittings indicate that the long range (linearly rising) part of the
potential
comes from scalar exchange. We included light quark pair creation in
such a model and found this coupling to decay channels incapacitated the model.
(See below.) On the other hand, models in which the quarkonium
potential as a whole is assumed vector exchange can account
well for    the  (spin-averaged) masses \cite{zamb};
`vector' models also account reasonably well for observed threshold
charmed and B meson production cross section data. Therefore, it may
be reasonable to assume there is some truth ion these `vector' models.
They are not inconsistent with the analyses of
Eichten and Feinberg \cite{eich-fein} and Gromes \cite{gromes} who obtain
in a $1/M^2$ expansion of QCD spin-dependent \QQ
interactions similar to those obtained in the Breit-Fermi reduction  of a
vector short range and
scalar long range potential.

\section{Including Dynamical Quarks}
\label{sec:dq}

Eichten et al. proposed a quarkonium potential model that includes
dynamical quarks nearly 20 years ago.
\cite{eichetal}  Their framework,
with various different dynamical assumptions, has been used by most
authors writing on this subject since then. Straightforward inclusion of light
quark pair
creation in a \QQ model requires quantum field theory and immediately
confronts us with a many body problem to solve. Eichten et al.,
hereafter referred to as the Cornell group,
simplified the problem to manageable proportions.
 Before going into details about model
predictions for charmed and
B meson productions, I would like to briefly outline their  framework.
Its structure incorporates fundamental properties of QCD,
namely confinement and asymptotic
freedom.
  Consider the  QCD vacuum polarization diagram shown below.
\begin{center}
\mbox{\epsfig{file=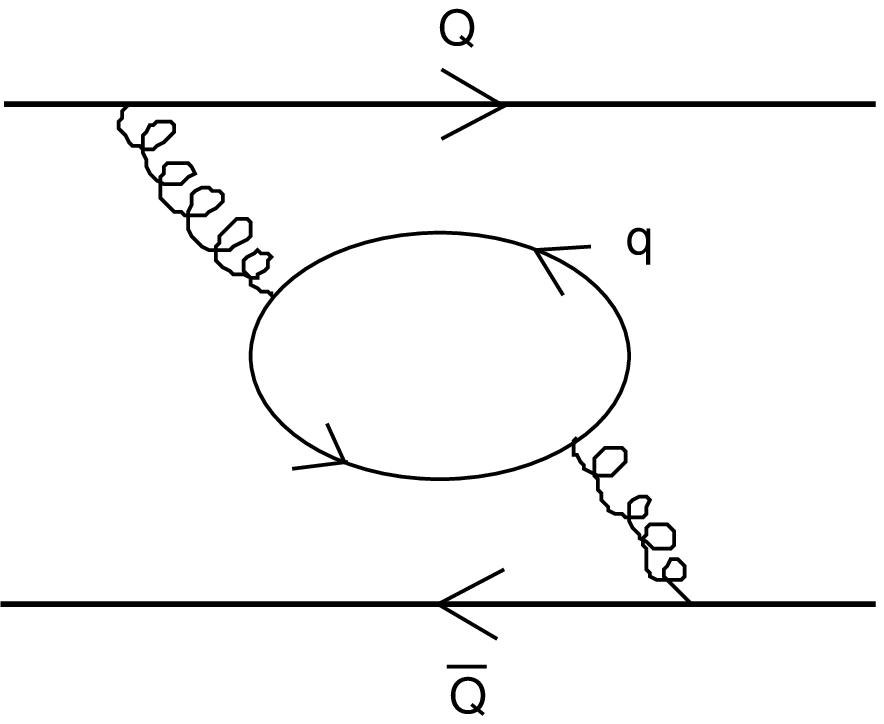,height=2.5cm}}
\end{center}
For loop momenta large compared to constitutent quark masses, this
diagram can be treated perturbatively and is taken into account using
the QCD running coupling constant.  However, diagrams with small
loop momenta should be treated
nonperturbatively.  For low momenta, higher order diagrams with multiple
gluon exchanges should be taken into account.
These were included by the Cornell group invoking
duality. They used a phenomenological field theoretic
model for \qq creation and then took the strong interactions
into account by replacing the \QqqQ
intermediate states  by \Qq and \qQ
mesons (bound by the same potential as binds \QQ). This reduces the many body
problem to a tractable
pair of two-body problems.

First
one solves a potential model for bound \QQ, \Qq and \qQ states. The
Cornell group used
\begin{equation} V(r) =
-{\kappa\over r} + C + {r\over a^2}\;. \label{eq:pot} \end{equation}
Then, as in
the Wigner-Weisskopf treatment of coupled channel problems
in nuclear physics, one solves a coupled channel
two-body problem that can be represented
by the Hamiltonian
\begin{eqnarray} H \: & = & \: \left( \begin{array}{cc} H_{Q
\overline{Q}} & h^\dagger \\ h & H_{Q \overline{q} q \overline{Q}} \end{array}
\; \; \right) \label{eq:ham} \end{eqnarray}
which operates in a space of bound \QQ states and bound \Qq and \qQ states.
In the Cornell model the \QQ states are eigenstates of
\begin{equation}
H_{Q\overline{Q}} = {p^2\over M_Q} + V + 2M_Q \;. \label{eq:one}
\end{equation}
and the \Qq and \qQ mesons eigenstates of the
corresponding Hamiltonian (with relativistic correction for the light
quark).  In the calculations, however, measured
(or to be measured) values of the \Qq and \qQ meson masses are used.
The piece denoted by $H_{Q \overline{q} q \overline{Q}}$ in (\ref{eq:ham})
is the kinetic energy operator for the two-meson states. The
off-diagonal piece $h$ couples the \QQ states to their two-body decay channels.
(Final state interactions are neglected; i.e., interactions between the
bound \Qq and \qQ mesons are neglected.)

To complete the framework described above, we need to put in a
 mechanism for pair creation. In the Cornell model, pairs
are  created by the same interaction that binds; i.e., by diagrams such as
\begin{center}
\mbox{\epsfig{file=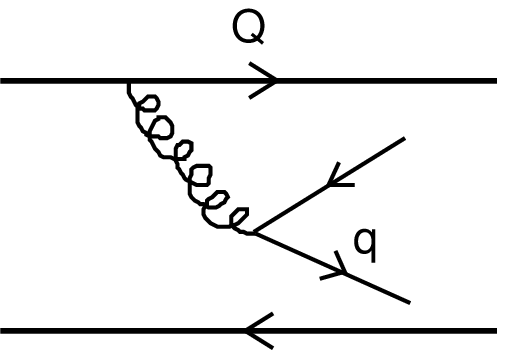,height=2.5cm}}
\end{center}
where the corkscrew line represents the propagator of an (instantaneous)
interaction whose Fourier transform is the quarkonium potential $V$.
The quark vertices, correspondingly, were taken as nonrelativistic
reduction of a vector exchange. One may vary these assumptions and
consider other dynamical models of quark pair creation. For example,
one could assume scalar rather than vector exchange.
As mentioned above, the fine structure in the mass spectra indicate
that the long distance part of the potential may be due to scalar
exchange. These two
assumptions give quite different results; a vector interaction creates
a pair with quantum numbers $1^{--}$ while a scalar creates a pair
with the quantum numbers of the vacuum. We can report here calculations
based upon various different assumptions about the dynamics
of \qq pair production.
They are tabulated in
Table II.

\begin{center}
\mbox{\epsfig{file=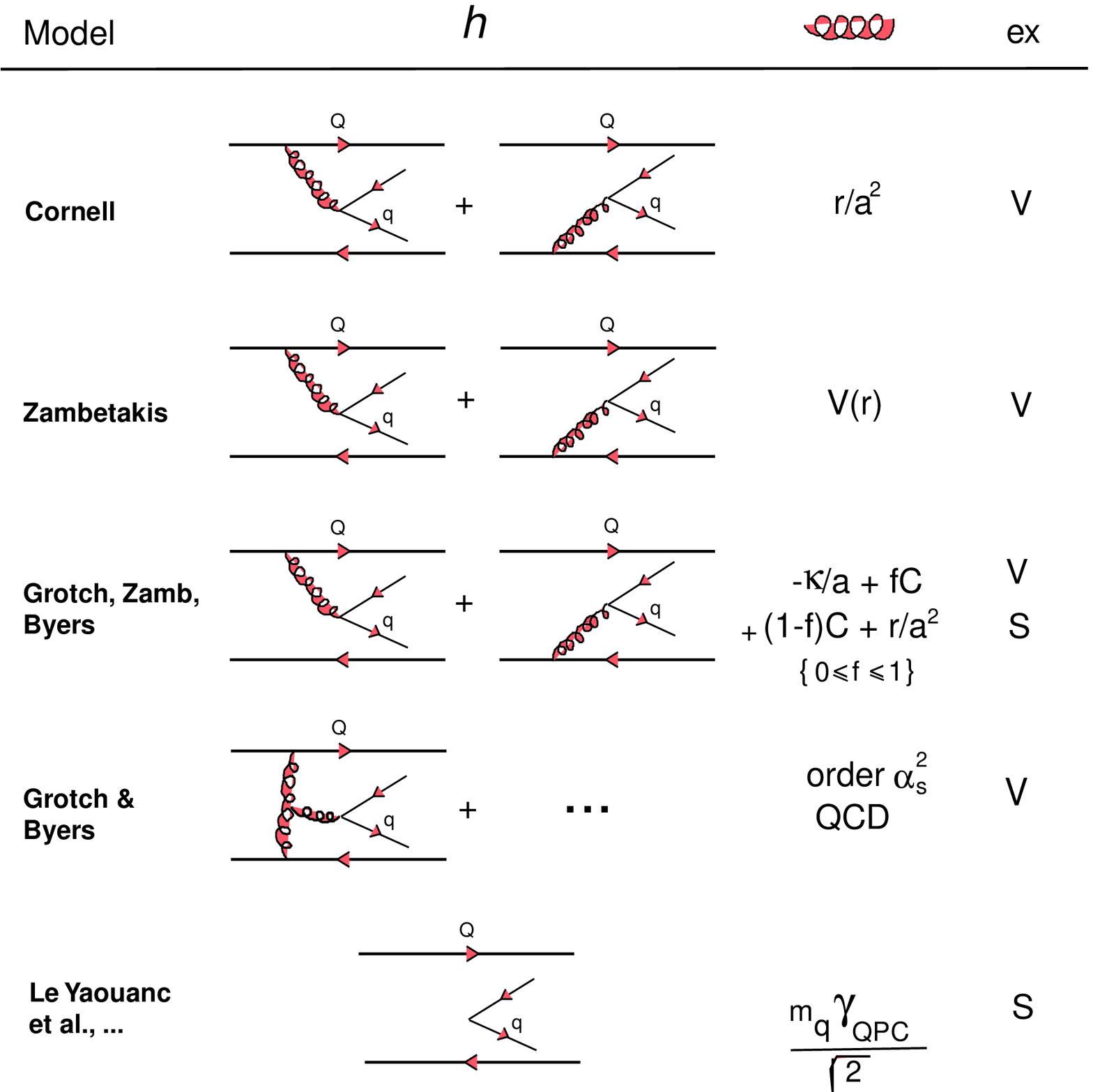,height=11cm}}
\end{center}

\begin{center} {\bf Table II.} \end{center}

The dynamical assumptions represented in Table II. are as follows. First,
the Cornell model: the original Cornell model neglected pair creation
from the Coulombic part of the potential and took pair creation only
from the long distance part. This is indicated in the third column.
In the fourth column is indicated whether the exchange is vector or
scalar. The second row represents Zambetakis' extension of the
original Cornell model to include the Coulombic part of the potential
in the pair creation matrix elements. \cite{zamb}
The third row represents our calculations with Howard Grotch in which
we took the Coulombic part of the potential to be vector exchange and
the long range part to be scalar. There is an ambiguity in implementing
this idea; namely where does short become long?  Quantitatively this
translates into the question of how to treat the constant term $C$
in (\ref{eq:pot}). For vector coupling, $C$ plays
no role; the pair creation matrix elements only depend on the gradient
of the potential.  However with scalar coupling the potential behaves
like a mass term and the pair creation
matrix elements depend on $V$ rather than ${\vec{\nabla}}
V$. Thus the constant $C$ can be significant. Fits to the data indicate
$C$ is large and negative; of the order of $-800$ MeV.\cite{zamb}
To take this ambiguiuty into account, we introduced another parameter $f$
which specifies which fraction of $C$ couples vectorially, and allowed it
to vary between 0 and 1.
The fourth row refers to calculations with Grotch of the
contribution of the triple gluon coupling graph.
We evaluated this graph in Coulomb gauge, and found its contributions
were not large; they are indeed of order $\alpha_s^2$. We also
studied the other QCD graphs of this order and found none seems to be
singular.  We therefore conclude that such effects are not at this stage
significant.  In the last row is represented the QPC model of Le Yaouanc
et al..\cite{leya} It is a model sometimes referred to as a flux tube
or string breaking model.  In it the \qq are produced with the quantum
numbers of the vacuum, and the pair creation matrix element is a constant
which can be included in calculations like the above by taking it
to be given by a constant potential $m_q \gamma_{QPC}/\sqrt{2}$. This model
gives interesting results which have been discussed in a number of papers;
see Ref.\cite{leya}.

In order to get the matrix elements of $h$ from the diagrams in Table II.,
one must sandwich them
between \QQ wave functions and
\Qq and \qQ. (Generally the \QQ wave functions are taken to be eigenstates of
(\ref{eq:one}), and for simplicity the \Qq and \qQ
wave functions are taken to be Gaussian fits to bound state
wave functions bound by the $r/a^2$
part of the potential.)

We report here the results of our studies of first four of these models.
Mainly what we have to report are our results for the Cornell and Zambetakis
models because (1) we found the order $\alpha_s^2$ graphs were not likely
to be large enough to be important; (2) the hybrid model in which the
long range part of the potential was considered scalar exchange was
unsuccessfull (explained below); and (3) both the Cornell and Zambetakis
models were able to account for available data - a large amount of data
on the narrow states below threshold and on production cross sections
for charmed and B mesons above threshold.  This is will be discussed
in detail in the last section of this paper.

We end this section with a brief explanation of why we think that the hybrid
model of short distance vector and long distance scalar exchange
is not a viable model.
Allowing
$f$ to vary between 0 and 1, we could not find values of the
parameters in the potential (\ref{eq:pot})
that allowed  a fit to the narrow state spectroscopic data.  The reason
for this is that in the region of $r$ where the quarkonium
wave functions are large $V$ changes sign, going from Coulombic
at short distances to linearly rising at large distances, and consequently,
 when it is coupled as a scalar, its contributions to matrix
elements of $h$ are relatively small differences of two large numbers.
Owing to this the matrix elements of $h$ vary erratically as one goes
from one of the low-lying quarkonium states to another;
consequently their contributions to
level shifts are unstable against  small variations of the parameters.

\section{Modification of NQM}
\label{sec:mod}

Before discussing our numerical results, it may be of interest to outline
here how the inclusion of dynamical quarks changes the NQM. This can
best be seen in a Hamiltonian framework in which the light quarks have
been integrated out and one has remaining an effective Hamiltonian
in the \QQ space.  This is
a peculiar Hamiltonian because (i) it is energy-dependent;
and (ii) for energies
above threshold for meson production, it develops an anti-hermitian part.
It is, however, quite interesting because it gives mass shifts and
configuration mixings and also gives production
cross sections for charmed and B mesons in \ee annihilation.

In the space spanned by eigenstates of (\ref{eq:one}), hereafter called
NQM states, the effective Hamiltonian is (see, e.g., Refs. \cite{zamb,cahn})

\begin{equation}
H_{eff}(W) = {\cal{M}}^{bare} + \Omega(W) \;. \label{eq:hef} \end{equation}
where ${\cal{M}}^{bare}$ is a diagonal matrix whose elements are the
eigenvalues of  (\ref{eq:one}). The $\Omega$ matrix  is second order
in $h$; the explicit expression for it is given below.
It is
perhaps useful to state some more of the properties of $H_{eff}$ here.
\begin{quote}
(1). The physical masses $M_N$ of the narrow quarkonium states
 are given by the solutions   to

\begin{equation}
 H_{eff}(M_N) \; a^N = \, M_N \, a^N  \label{eq:eig}
\end{equation}
where $a^N$ denotes a state vector in the space spanned by NQM states.
A physical state  ${\bf{\Psi_N}}$ may be expanded as
\begin{equation}
{\bf \Psi_N} = \sum_{i} a_i^N\psi_i + \sum_{n} b_n^N \phi_n \;. \label{eq:Psi}
\end{equation}
where $\psi_i$ are NQM statesand $\phi_n$ are continuum
two-meson states.  The $a_i^N$ are the components of the
`eigenvectors' $a^N$ in (\ref{eq:eig}).  The problem (\ref{eq:eig}) is
not exactly an eigenvalue problem.  The dependence on $M_N$ is highly
nonlinear.  Nevertheless  solutions can be found.\cite{zamb}\\

(2). $H_{eff}(W)$ is not hermitian when $W$ is above the threshold of the
lowest mass two-meson continuum.   The quarkonium
states whose masses are above threshold are poles in the complex W plane.
They are poles of the effective propagator  ${\cal{G}}(W)$, where

\begin{equation}
{\cal{G}}(W)\; = \left(W \, - \, H_{eff}(W)\right)^{-1} \;.  \label{eq:gee}
\end{equation}
For $W$ above threshold, $\Omega$ has the structure
\begin{equation}
\Omega(W) \, = \, R(W)\, - \, i \Gamma(W)/2  \label{eq:nonher}
\end{equation}
with $R$ and $ \Gamma $ real symmetric matrices. Unitarity requires
the diagonal elements of $\Gamma $ to be positive.\\
\end{quote}
The matrix elements of $\Omega(W)$ are given by
 are given by
\begin{equation}  \Omega_{ji}(W) = \sum_n {(\psi_j, \,
h^{\dagger}\phi_n) (\phi_n, \, h\psi_i)\over(W - E_n + i\epsilon)} \;.
\label{eq:omeg}
\end{equation}
The sum over $n$ denotes sum over channels and
integration over channel phase space; $E_n$ is the channel energy.
The channels are
labeled by $f$ which specifies channel spins, flavor, masses, etc..
Because of this sum over channels,
$\Omega$ is a sum of matrices
\begin{equation}
\Omega(W) \,= \sum_f \,\Omega^{(f)}(W) \;.  \label{eq:omegf}
\end{equation}
Each matrix element of $\Omega^{(f)}$ is an integral
 which becomes singular when  $W \geq m_1 + m_2$ where $m_1$ and $m_2$
are the masses of the mesons in channel $f$. These singular integrals
are evaluated in the usual way taking $W$ in the upper half plane and
then the boundary value on the real axis from above.

For $W$ above flavor threshold, charmed and B meson
production cross sections in \ee annihilation may be calculated
as follows. The cross section ratio
\begin{equation}
\Delta R^f =
{\sigma(e^+\,e^- \, \rightarrow \, f)\over\sigma(e^+\,e^- \, \rightarrow
\, \mu^+\,\mu^-)}\;, \label{eq:delr}  \end{equation}
in one photon
approximation,
is given by the dispersive part of ${\cal{G}}(W)$.
With a high speed computer  and $H_{eff}$ in matrix form (see below),
${\cal{G}}$ is easily calculated by matrix
inversion.
There is no coupling between the \cc and \bb subspaces, so one deals separately
with these two subspaces and
the calculations for the production cross sections of charmed and B mesons
are done independently.
The charm contribution to the cross section ratio $\Delta R_c$
can be expressed
as a trace in the \cc subspace; viz.,

\begin{equation}
\Delta R_c  = -{72 \pi \over W^2}\,  e_c^2\;
{\bf{Tr}} (\,{\cal W}\, {({\cal G} -{\cal G}^
{\dag})\over 2i}\,)_{c\overline{c}}
\label{eq:delrc}
\end{equation}
where $e_c = 2/3$,  and the matrix $e_c^2~{\cal W}$ is bilinear in
the \cc-photon coupling.
Nonrelativistically, matrix elements of ${\cal{W}}$
are given by \cc NQM wave functions; viz.,
\begin{equation}
{\cal{W }}_{ij} = \,\psi_i (0) \psi_j (0) \;. \label{eq:calw}  \end{equation}
(Relativistic corrections are important here, particularly for charm; see Ref.
\cite{novikov}.) In their original papers, the Cornell group showed that

\begin{equation}
{\cal G}\, -\,{\cal G}^{\dag}\; =  {\cal G}^{\dag} ( \Omega - \Omega^{\dag})\,
{\cal G} \,. \label{eq:imomeg}  \end{equation}
Thus, from (\ref{eq:delrc}), (\ref{eq:imomeg}), and (\ref{eq:omegf}), one sees
that the
exclusive production cross section for channel $f$ is
\begin{equation}
\Delta R_c^{(f)}  = -{72 \pi \over W^2}\  e_c^2\
{\bf{Tr}} (\;{\cal W}~{\cal G}^
{\dag}~Im  \Omega^{(f)}~ {\cal G}\;)_{c\overline{c}}\,.
\label{eq:delrf}
\end{equation}
Similarly, one calculates production cross sections for B mesons.
{}From the above it is clear that this is a nonperturbative treatment
of dynamical quarks.

With high
speed computers it is straightforward  to solve, as outlined above,
for both the narrow states below threshold
and the (resonant) cross sections above threshold. One first solves
(\ref{eq:one}) for the NQM wave functions and masses.
The matrix elements of the $\Omega$ matrix and
(\ref{eq:hef}) can then be explicitly evaluated.
These matrix elements are discrete because, owing to confinement,
the entire \QQ state space  is spanned by a discrete set of
eigenfunctions.  With these matrices one can calculate all of the
above quantities.
Most computers are now able to
work with matrices of (almost) arbitrarily high dimension.  However,
since we are interested only in the low lying quarkonium states,
 we do not need to calculate matrices of
high dimension. The highly excited NQM states make
negligible contributions.

On the other hand, another approximation involved
in a numerical evaluation of $\Omega$(W)  may
cause significant error in predicted
masses of quarkonium resonances in the continuum.
In calculating $\Omega$(W) one generally approximates the
infinite sum over channels $f$  by
a finite sum neglecting all channels with thresholds
greater than some minimum energy $E_{min}$ which is greater than
the maximum value of W for which one is calculating
the $\Omega$ matrix. Though the contribution
from any one of these neglected
channels is small, their cumulative effect can  be
appreciable because all these channels contribute coherently
(negatively) to the diagonal elements of Re $\Omega$; c.f., Eq.
(\ref{eq:omeg}). (Cancellations are likely in contributions to
the off-diagonal elements owing to random phases.)
  These cumulative mass shifts are small for W
substantially below $E_{min}$; however, they may become appreciable when
W comes close to $E_{min}$. \footnote[1]{In our calculations we included
 only the channels with ground state pseudoscalar and
vector mesons.
This is a good approximation for the narrow states below flavor
threshold;   however, for values of W in the continuum
we found it necessary, as did the Cornell group,
to compensate for the neglect of excited meson
decay channels by putting in a phenomenological mass shift
to fit the observed resonance energies
\cite{eichetal,by-eich} .}

\section{Production Cross Section Calculations and Comparison with Experimental
Data}

In this section we report our results on
threshold production
cross sections, inclusive and exclusive, for charmed and B mesons
in \ee annihilation and compare them with available data. There are two
sets of results; one obtained from the original Cornell model and the
other from Zambetakis' extension in which pair creation occurs at short
distances also.
The available data are mainly
inclusive cross section measurements.  The agreement with data is about the
same for both although it differs in detail in the two cases.  The two models
give significantly different results for exclusive channel cross sections.
Measurement of these may distinguish between them. The degree of agreement
of these model predictions with data is, in our view, significant because
the  calculated
cross sections are obtained without adjustment of parameters or
introduction of additional parameters. The parameters are
the parameters of the potential $\kappa,~C,$ and
$1/a^2$. We calculated the masses of the narrow charmonium states using
(\ref{eq:eig}) and determined these three parameters by fitting the
observed  masses of
$J/\psi,~\chi_{cog},~{\rm{and}}~ \psi'$.\setcounter{footnote}{1}
\footnote{After a  detailed and extensive analysis, the Cornell group
used the following constitutent quark masses
 $M_c$ = 1.84 GeV, $M_b$ = 5.17 GeV, and for light quarks,
 $m_u = m_d $ = 335 MeV, and $m_s$ = 450 MeV. We used the same values in our
calculations.}
Parameter values  which fit these data are\footnote{The slope of the
linear potential $1/a^2$ is significantly greater  in Zambetakis' model because
the short distance pair creation increases the coupling of \cc
to inelastic channels This stronger coupling to inelastic
channels increases
 the separation of charmonium energy levels, and requires increased
strength of the confining potential   to bring the
 mass differences back to their observed values.}
\begin{center}
\mbox{\begin{tabular}{c||c|c|c|c}
model &  $1/a^2$ in GeV$^2$  &  $C$ in GeV &
$\kappa(c\overline{c})$\\
 \hline\hline
Zambetakis &  0.31  &  -0.97 &  0.49  \\
\hline
Cornell  & 0.22 & -0.85 & 0.52
\end{tabular}}
\end{center}
Using these parameters, we calculated (spin averaged) $\Upsilon$ masses,
charmonium and $\Upsilon$ leptonic widths, etc.
and found good agreement
with measured values \cite{zamb}.\footnote{For the $\Upsilon$ masses and
widths,
and for  B meson production cross sections, the QCD logarithmic decrease of
$\alpha_s$  was taken into account by using for $\kappa(b{\overline{b}})$
the values 0.46 and 0.48 for the Zambetakis and Cornell models, respectively.}
 Then we calculated
 charmed and B meson production cross sections above flavor threshold
evaluating (\ref{eq:delrf}) with the same parameters. We calculated charmed
meson productionfrom threshold
to 4.5 GeV, and  for B from threshold to 11.1 GeV.
In the numerical calculations
we truncated the \cc state space  after the 4S and 2D
 NQM states, as in the original Cornell calculations. We
found that including more states did not significantly alter the results.
For B meson production, the $\Upsilon$(4S) is the first \bb resonance above
threshold; we truncated after the 7S and 4D NQM states. Results
were stable against increaseing the number of such states.
However, we also found that the results changed little when we omitted
the D-states  and only changed noticeably near 11 GeV
when we omitted the 7S state. The results
presented here were obtained truncating \bb states at the 6S.
The reason it was not necessary to include D-states is that owing to the
heavier b quark mass, the
S-D mixing  and photon-D state coupling are smaller than in the charm case.

\subsection{ Charmed meson production.}

To understand the complicated energy dependence of the
inclusive charm production cross section above threshold (3.73 GeV), it
is useful to consider the singularities of the 1$^{--}$ \cc propagator
${\cal{G}}_{c\overline{c}}(W)$ in the complex energy plane; ${\cal{G}}(W)$ is
given by
(\ref{eq:gee}).  The singularities are cuts along the real axis and
poles in the lower half plane.  There is a cut for each open channel
with a branch point at its threshold.  The poles correspond to resonances in
the
cross section. The branch points and poles are shown in Fig. 1 where the
positions of the singularities are determined by the experimental values of the
masses and  widths of the indicated states \cite{pdg}; the
distance  below the real axis of each  pole is the measured total half-width.
 The separations of the branch points and
positions of the
poles are to scale.
\begin{center}
\mbox{\epsfig{file=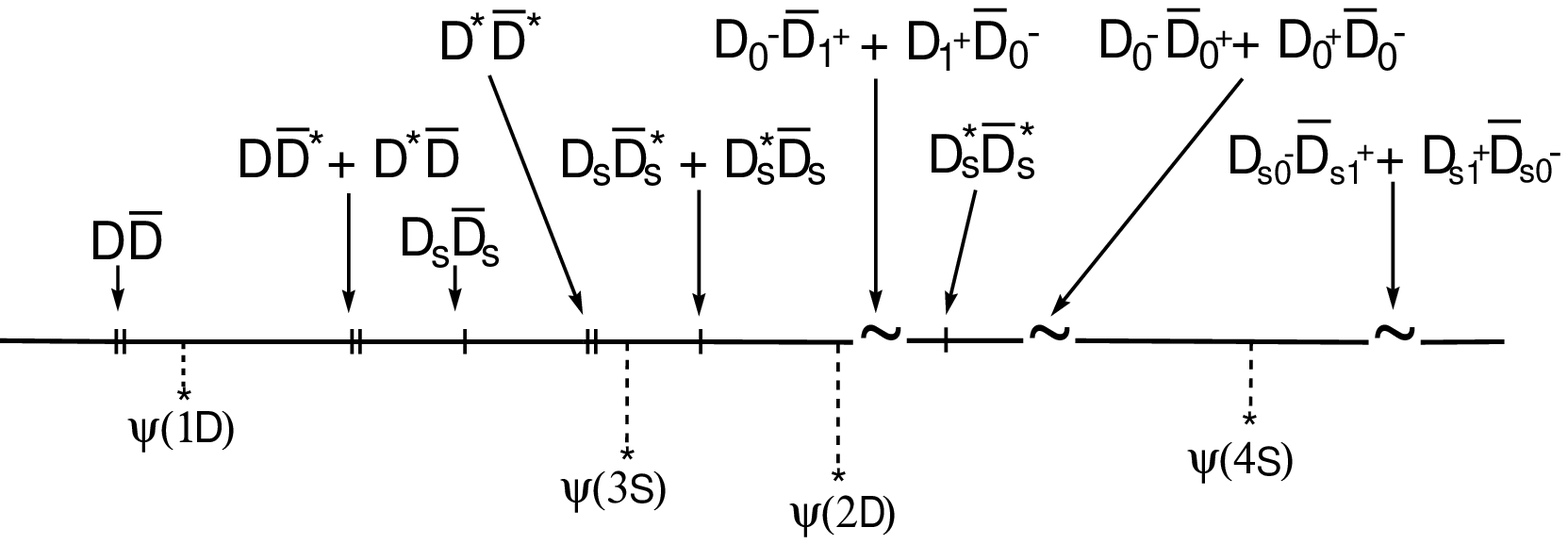,height=4.0cm}}
 \end{center}
\begin{center} {\bf{Figure 1.}}   \end{center}
The corresponding cross section measurements and theoretical curves
for R are shown in Fig. 2.  The first resonance above threshold is
understood, from the point of view of the Cornell model, as the $\psi(1D)$
though it is not a pure state but has admixtures of other nearby $1^{--}$
S and D states. Though this is a relatively
narrow resonance  as indicated by its proximity to the real axis
in Fig. 1, it is not as prominent as the
S-state peaks,
 $\psi(4040)$ and $\psi(4415)$, owing to the fact that nonrelativistically
D states do not couple directly to the photon.  As it is primarily a D-state,
the $\psi(3770)$ owes its presence in an \ee annihilation cross section to
the fact that it has some configuration mixing with nearby S states (
particularly the 2S) and a direct coupling to the photon in order
a $(v/c)^2$.  Similarly the $\psi(2D)$ appears in the cross section as
a shoulder on the $\psi(3S)$ or $\psi(4040)$ peak owing to its admixture of
3S-state and direct coupling to photon.
\begin{center}
\mbox{\epsfig{file=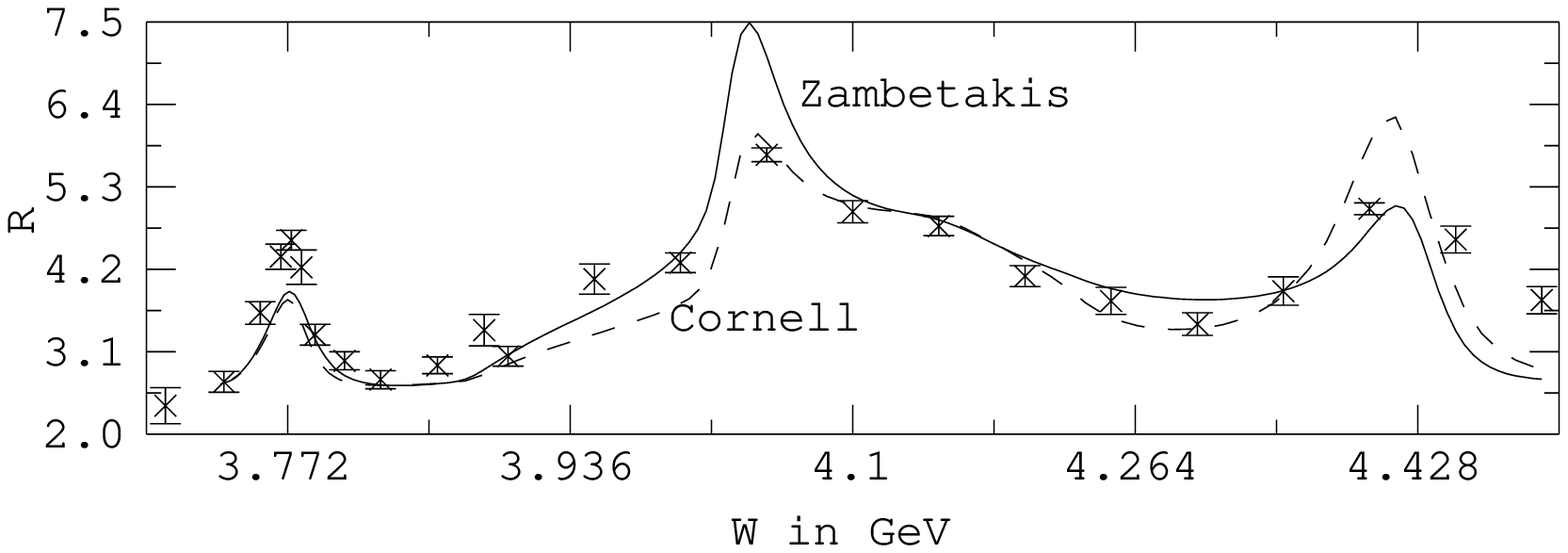,height=4.5cm}}
\end{center}
\begin{center} {\bf{Figure 2.}} \end{center}
The data points in Fig. 2 are from the {\it{Review of Particle Properties}}.
\cite{pdg} The curves are calculated using the Zambetakis and Cornell models
 with the contribution to R from  other than charmed hadrons
 taken from the data to be 2.65; i.e., the curves in Fig. 2
are $\Delta R_c(W) + 2.65$.\footnote{ A phenomenological
mass shift matrix was used as compensation for the neglect
of excited meson decay channels; we only included
the ground state psuedoscalar and vector meson decay channels
in our calculations, c.f., section V.}
What is remarkable about these models is the degree to which the
energy dependence and overall normalization of
the calculated curves are in agreement with the data.
Neither model gives a perfect fit to the data;
but considering the simplicity of the models, and the complexity of the
physics, it seems remarkable that the models fit the data as well as they do.
The Cornell model seems to fit the data in the region of the $\psi(4040)$
better and the Zambetakis model seems to do better in the region of the
$\psi(4415)$. It is
difficult on the basis of these inclusive cross section data to
say if one or the other of these models is closer to being correct.

The models are more easily distinguished in their predictions for the
exclusive cross sections.  In Figs.
3 and 4 are shown the individual channel contributions to $\Delta R_c$.
\begin{center}
\mbox{\epsfig{file=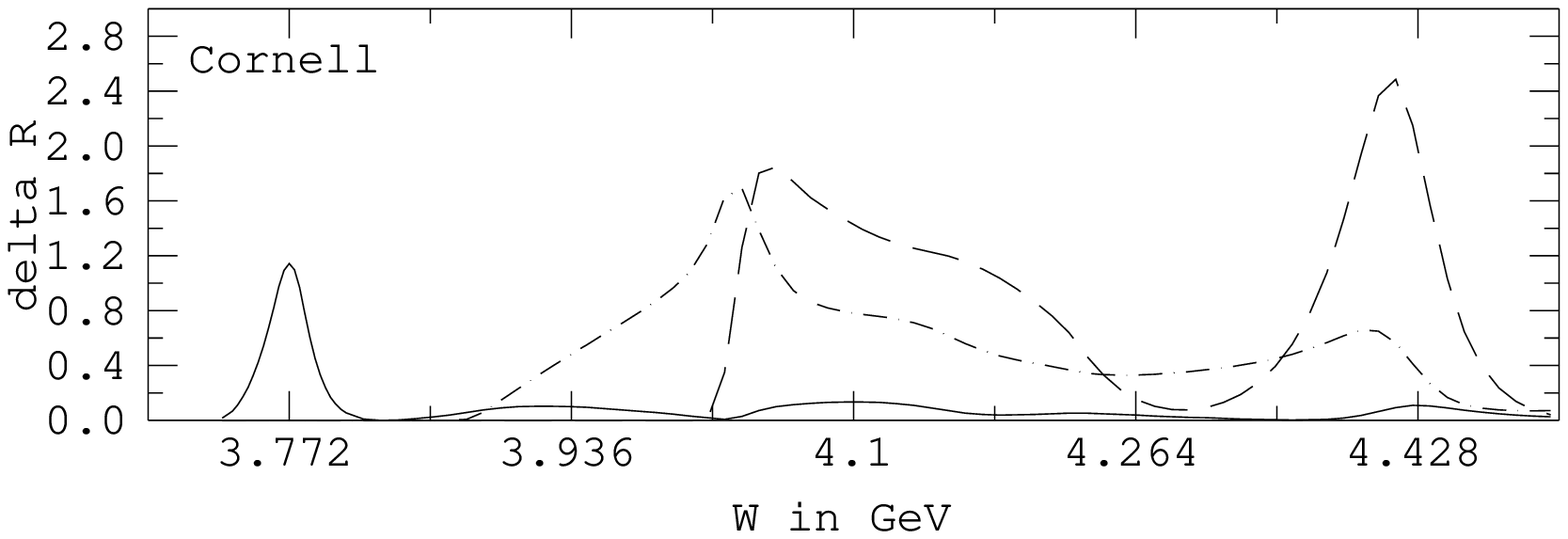,height=4.5cm}}
\end{center}
\begin{center} {\bf{Figure 3.}} \end{center}
The full curve is the contribution to $R_c$ from $D\overline{D}$
channels, both charged and neutral;  the dot-dashed curve
from $D\overline{D}^*~+~D^*\overline{D}$; and the
dashed curve from $D^*\overline{D}^*$.
\begin{center}
\mbox{\epsfig{file=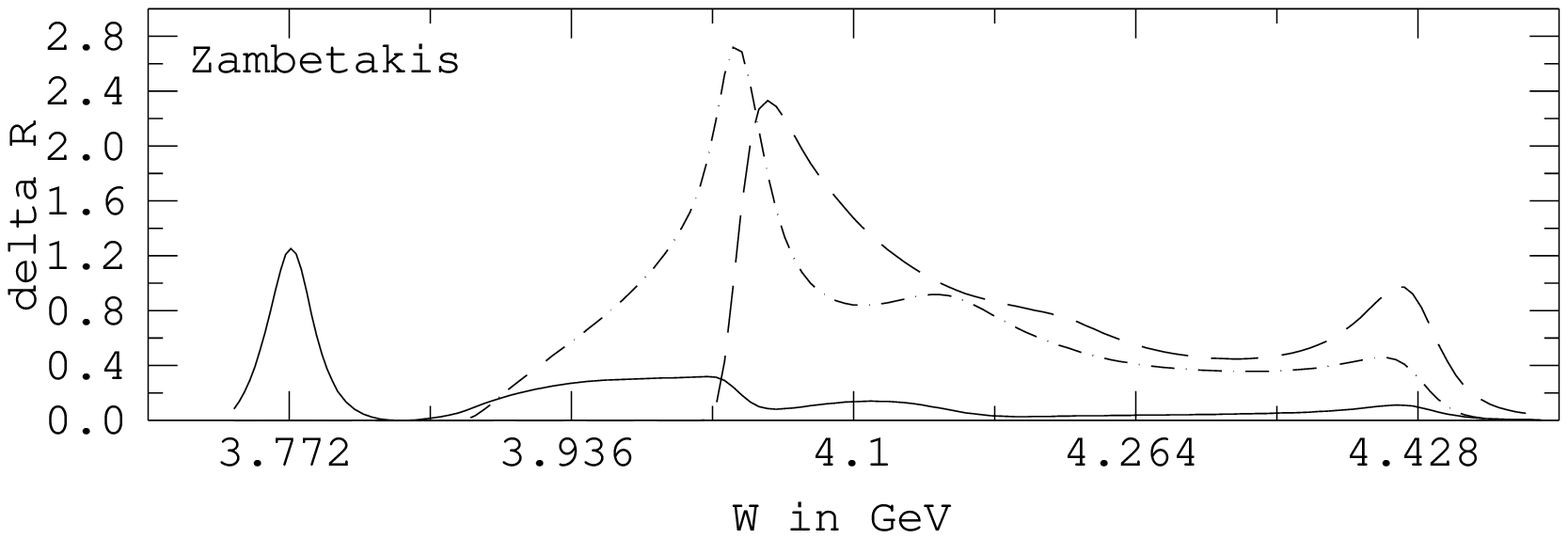,height=4.5cm}}
\end{center}
\begin{center} {\bf{Figure 4.}} \end{center}
The cross sections for $D_s$ and $D_s^*$ production
are roughly 1/3 of those for non-strange
D mesons. These are shown in Fig. 5.
\begin{center}
\mbox{\epsfig{file=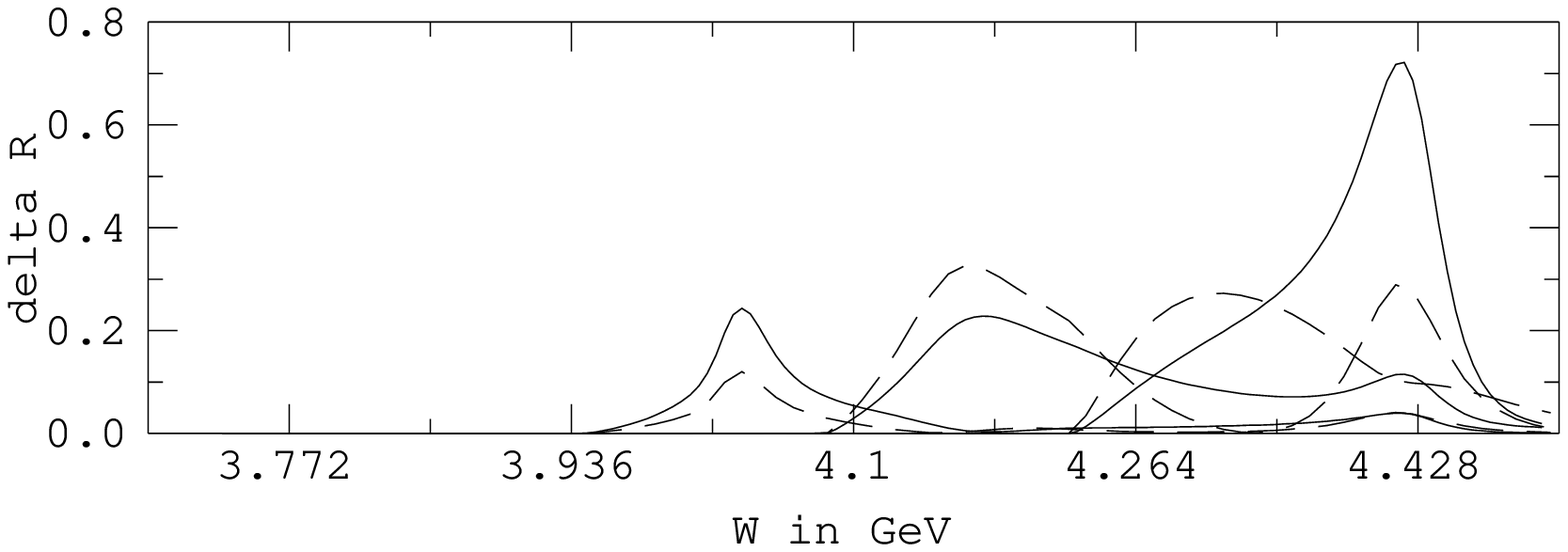,height=4.5cm}}
\end{center}
\begin{center} {\bf{Figure 5.}} \end{center}
In Fig. 5 the full curves are Cornell model predictions for $D_s
\overline{D}_s$,
$D_s \overline{D}_s^* +  D_s^* \overline{D}_s$,  $D_s^* \overline{D}_s^*$.,
and the dashed curves Zambetakis model predictions. The curves for the
different
channels can be distinguished by the fact that they start at different
thresholds.
Note the scale
change. This suppression of the strange mesons
is natural to these models. It comes from the $m_s$ being greater than
$m_u$. The suppression is due, essentially, to two effects: (i) the pair
creation amplitude is inversely proportional to the
light quark mass, and (ii) the strange quark meson states are smaller
(more tightly bound).

\subsection{B meson production cross sections.}

The singularity structure of ${\cal{G}}(W)_{b\overline{b}}$ relevant for
our purpose is shown in Fig. 6. It appears simpler than Fig. 1 because
the $B^*-B$ mass difference is smaller than $D^*-D$ and because we
have indicated only the S-state resonances. In the $\Upsilon$ system
both S-D  mixing and direct photon-D-state coupling are small. These are
effects of order $(v/c)^2$ and therefore
smaller than in charmonium. Consequently
D-state contributions are not significant in
\ee annihilation cross sections.
 \begin{center}
\mbox{\epsfig{file=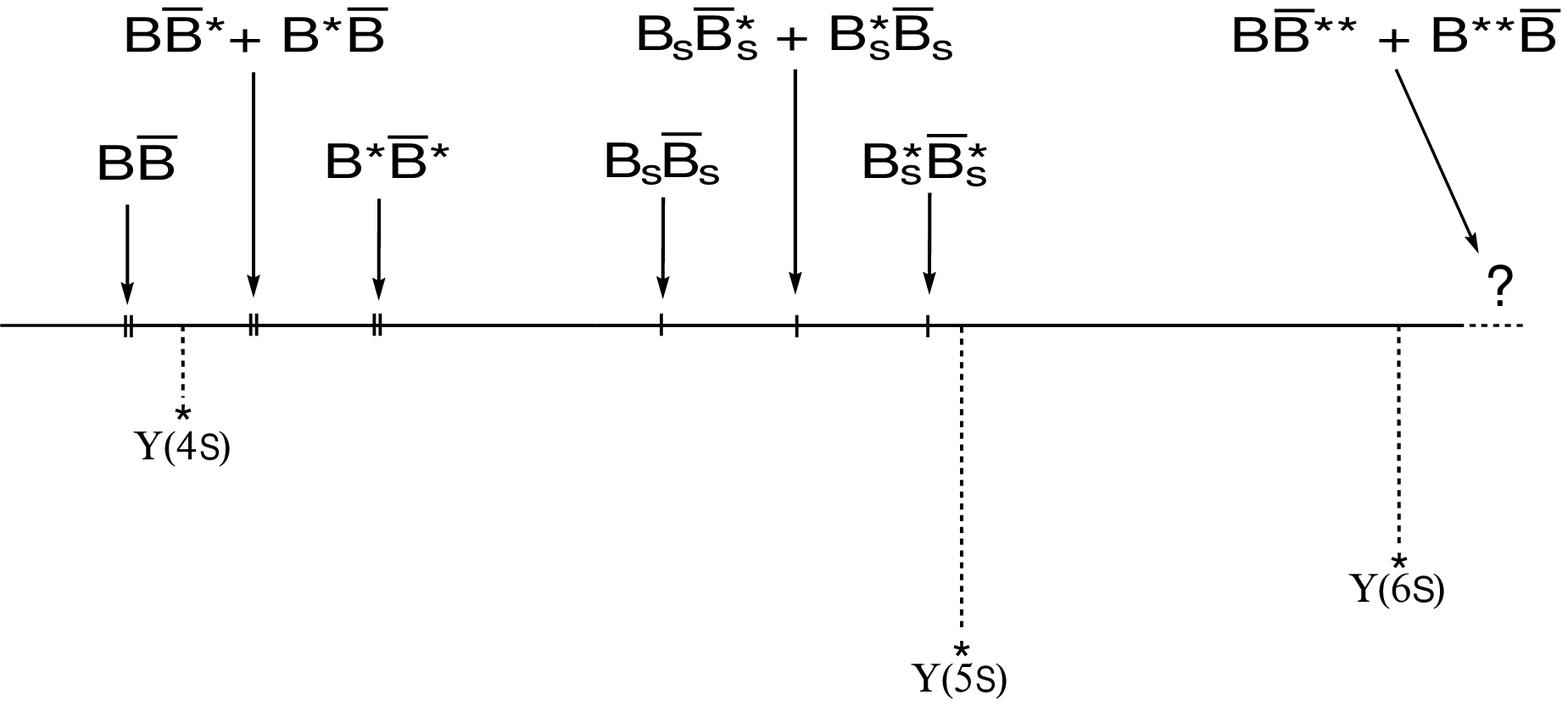,height=4cm}}
\end{center}
\begin{center} {\bf{Figure 6.}} \end{center}
Corresponding to this in Fig. 7, we show the inclusive
B meson production cross section data\cite{besson}
along with the calculated curves.
 \begin{center}
\mbox{\epsfig{file=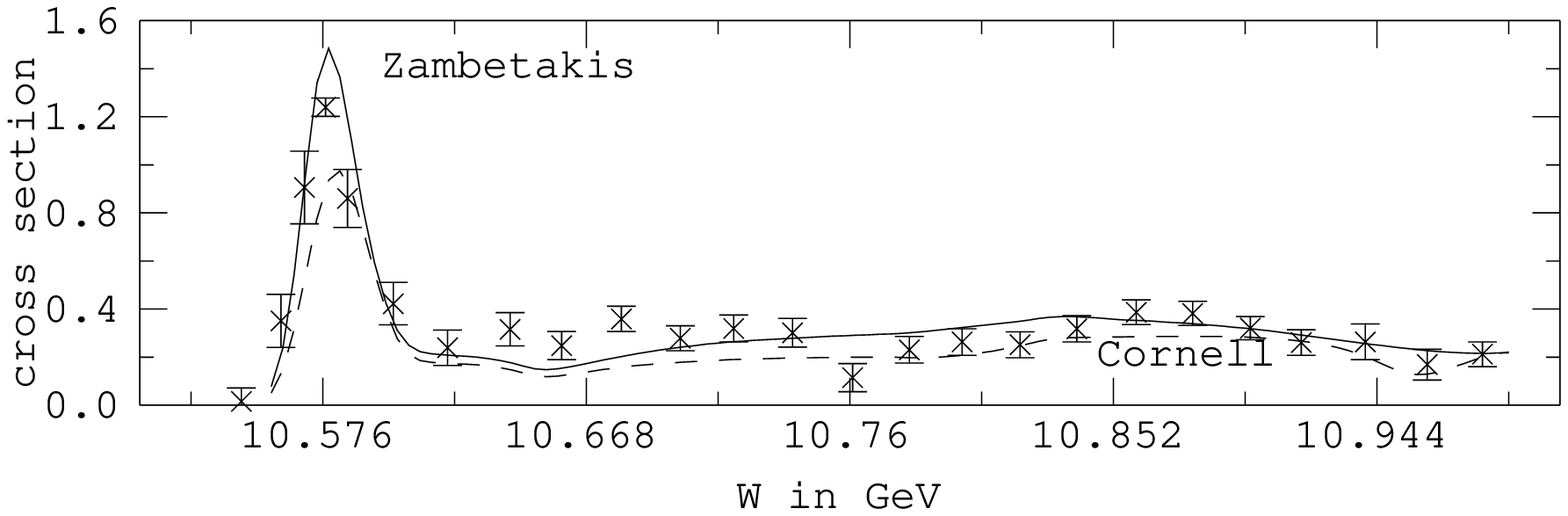,height=4.5cm}}
\end{center}
\begin{center} {\bf{Figure 7.}} \end{center}
Radiative  and beam spreading
corrections have been applied to the theoretical curves in Fig. 7
so that they can be
compared with the data which are uncorrected for radiative corrections and
beam spread.\footnote{We thank Dave Besson, Persis Drell and Elliot Chu  for
applying the radiative and beam spread corrections to the calculated cross
sections.}

Owing to the fact that the $\Upsilon$(4S) can only decay to $B\overline{B}$,
it is a relatively narrow resonance and prominent in the data.  The
$\Upsilon$(5S), on the other hand, can decay to all nine ground state
psuedoscalar and vector mesons and
is relatively broad. In Fig. 6 its position is far from the real axis
and  relative to its width, its mass is near many sharp thresholds;
the mass and width are
10.865 GeV and 110 MeV, respectively.   Because of these complications,
the $\Upsilon$(5S) does not have a normal Breit-Wigner shape and is hard
to see in the cross sections.  The $\Upsilon$(6S) mass is quoted
in Ref. \cite{pdg} as 11.019 GeV. We have included in Fig. 7
data only out to 10.99 GeV because we think that our calculated values in the
region of the 6S are unreliable owing to the fact that we have not included
channels with B mesons in excited states -
either with $\ell \neq 0$ or excited $\ell = 0$ mesons; and these are likely to
have thresholds in the vicinity of the $\Upsilon(6S)$.

The curves in Fig. 7 have been calculated without any adjustment of the
parameters of the model. It is, therefore, in our opinion remarkable that
they agree with the data as well as they do. As indicated at the end of
section V, a mass shift matrix has been used to fix the
peak values of the cross section at 10.580 GeV and 10.865 GeV.  Aside
from this, the energy dependence and the overall normalization of the
calculated curves are determined by the models.
The Zambetakis model (full
curve) appears to fit the data around the $\Upsilon$(4S) better than the
Cornell model (dashed curve).  However, in the region of the  $\Upsilon$(5S)
the Cornell model may be a better fit.

As in the charm case, the energy dependence of the exclusive cross sections
are quite different for the two models.  The contributions to the cross
section ratio R from the various channels
are shown in Figs. 8 and 9;
the full curves are for the $B\overline{B}$ channels, neutral plus charged;
the dot-dashed curves are for $B\overline{B}^* + \overline{B} B^*$; and the
dashed curves are for $B^*\overline{B}^*$.  Notice that at the higher
energies where these curves tend to flatten out they go over toward the
proportion 1:4:7 which are the spin ratios, discussed in Ref. \cite{eichetal},
  for  $B\overline{B}$ ,
 $B\overline{B}^* + \overline{B} B^*$, and  $B^*\overline{B}^*$ , respectively.
 In Figs. 8 and 9 we omitted the $\Upsilon$(4S) resonance
region because the cross section there is so much bigger than it is in the
rest of the range.
\begin{center}
\mbox{\epsfig{file=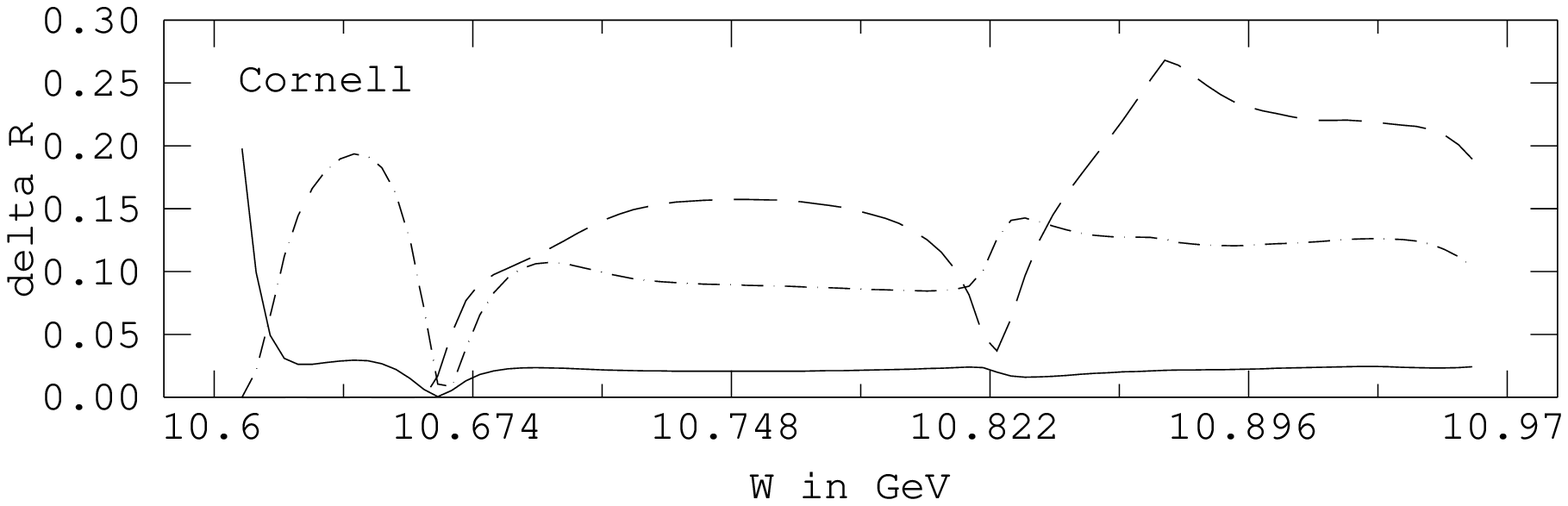,height=4.5cm}}
\end{center}
\begin{center} {\bf{Figure 8.}} \end{center}

\begin{center}
\mbox{\epsfig{file=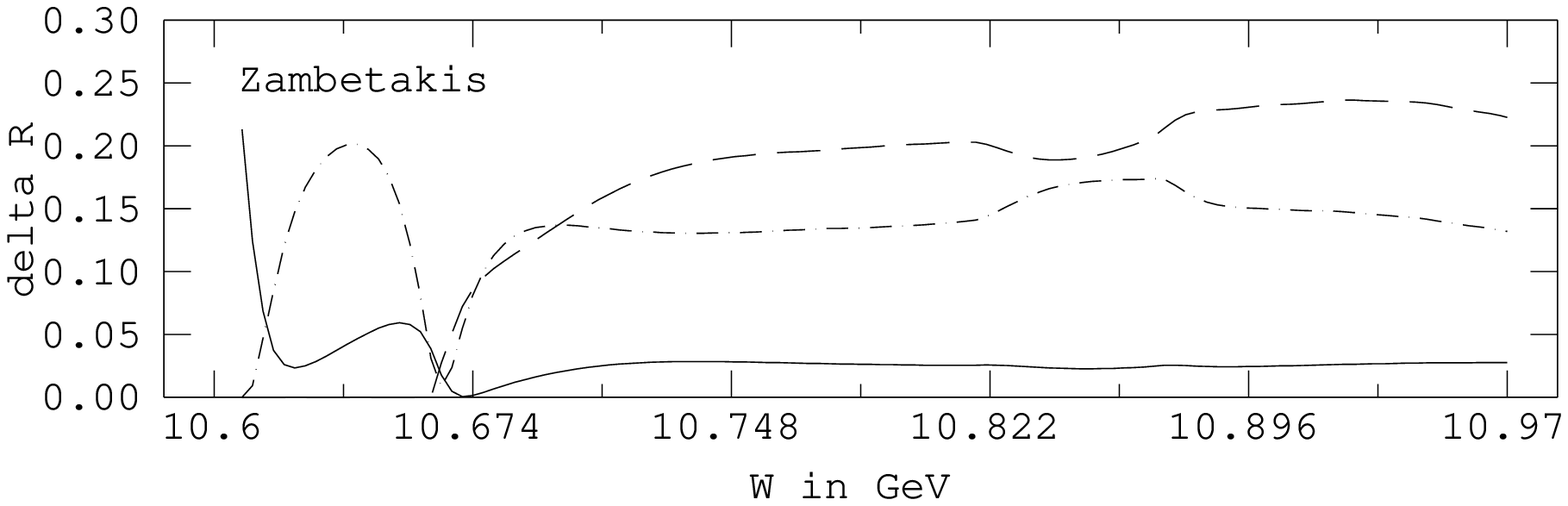,height=4.5cm}}
\end{center}
\begin{center} {\bf{Figure 9.}} \end{center}
Finally in Fig. 10 we show the contributions to R from $B_s$ channels;
the full curves are Cornell model predictions and dashed curves
Zambetakis model. The curves for the various channels can be distinguished
by their thresholds.
\begin{center}
\mbox{\epsfig{file=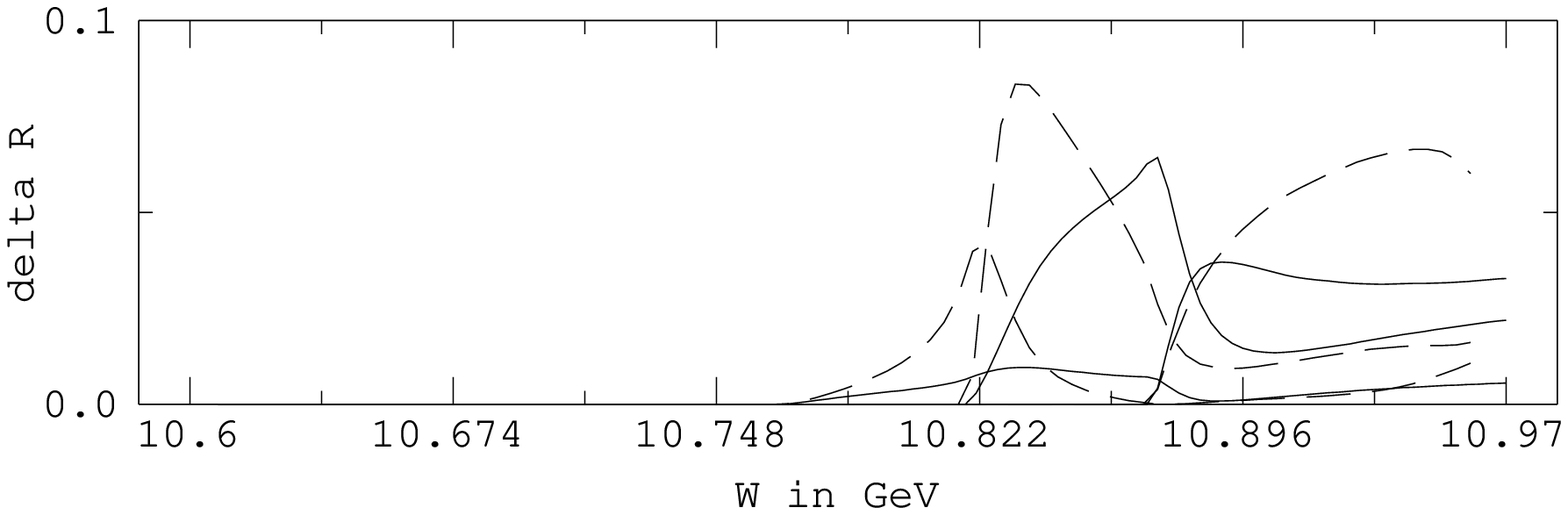,height=4.5cm}}
\end{center}
\begin{center} {\bf{Figure 10.}} \end{center}
One interesting feature of the curves in Fig. 10 is that both models predict
 relatively copious
$B_s\overline{B}^*_s + \overline{B}_s B^*_s$ production in the
$\Upsilon$(5S) resonance region.  (These predictions are sensitive to
the $B_s$ mass and the $B^*_s - B_s$ mass difference.)
 As in the charm case, however, these models predict that
the strange mesons are produced at a level of about 1/3 that of the
nonstrange  mesons.  Note the change of scale in Fig. 10.

\section{Conclusions}

The fact that both the
the Cornell and Zambetakis models, as simplistic as they are, can account
very well for charmonium and $\Upsilon$ narrow state spectroscopy and,
  without adjustment of parameters,
give production cross sections which agree as well as they do with
the data  leads us to conclude that they merit further
study. There are two directions for this.  Experimentally, it would be
interesting to compare calculated values for exclusive cross sections
with measured values. From theoretical point of view, it may be
worthwhile to evaluate taking predicted configuration mixing into account
the allowed and forbidden M1 transition rates in charmonium and $\Upsilon$
and compare those with measured values.

\acknowledgements

It is a pleasure to acknowledge the contributions to the
computational side of this  work from
Stanley Cohen and staff of Speakeasy Corporation, and from the
UCLA Office of Academic Computing. This work was supported in part
by the Department of Energy grant FG03-91ER40662.


\begin{references}
\bibitem{qq} In this paper
we will use $Q$ for $c$ or $b$ and $q$ for $u$, $d$, or $s$.

\bibitem{eichetal} E. Eichten, K. Gottfried, N. Kinoshita, K.D. Lane, and
H.- M.Yan, Phys. Rev. D17, 3090 (1978); D21, 313(E)(1980); and
Phys. Rev. D21, 203, (1980).
\bibitem{eich-fein} Estia Eichten and Frank Feinberg, Phys. Rev. D23, 2734
(1981).
\bibitem{gromes}  Dieter Gromes, Z.Phys. C26, 401 (1984). See also Phys. Lett.
202B, 262 (1988).
\bibitem{zamb} V. Zambetakis, UCLA Ph.D. thesis (1985);
available as UCLA research report UCLA/85/TEP/2.
\bibitem{grot}  H. Grotch, V.Zambetakis and N. Byers. (unpublished)
\bibitem{mccl} Richard Lee McClary, UCLA thesis (1982);
N. Byers and R. L. McClary, Phys. Rev. D28, 1692(1983)
\bibitem{bram} N. Brambilla and G. M. Prosperi, Phys. Lett. B236,69 (1990).
\bibitem{cahn} T. Appelquist, R. M. Barnett, K. Lane in {\it {\ee Annihilation:
New Quarks and Leptons}}, Benjamin/Cummings pub., R. N. Cahn, ed..

\bibitem{novikov} V. A. Novikov et al., Phys. Rev. Lett. 38, 626 (1977)
These authors pointed out that there is a
a $(v/c)^2$ correction which  gives
direct  coupling of the photon to  $^3 D_1$  states.
 There is in addition a one gluon exchange correction to
the quark-photon vertex which the  Cornell group includes
by multiplying the $\ell = 0$ subspace of ${\cal{W}}$ by
$(1-4\kappa/\pi)$.



\bibitem{by-eich} N. Byers and E. Eichten {\it{Heavy Flavour Production near
Threshold in \ee
Annihilation}} (unpublished). N. Byers and E. Eichten, Nucl. Phys. B(Proc.
Suppl.) 16, 281 (1990).
\bibitem{leya} Le Yaouanc, L. Olivier, O. Pine,
and J. C. Raynal, Phys, Rev.
D $\underline{8}$, 2223 (1973);
See also  S. Ono, A. I. Sanda,
and N. A. Tornqvist,Phys. Rev. D $\underline{35}$, 907 (1987)
and references cited therein.Studies of B production in this model
have also been made by Martin and Ng; cf.,
A. D. Martin and C.-K. Ng,preprint DTP/88/6. Fits to the production cross
sections are made with parameters in addition to those in the \QQ potential
and the quark pair creation strength $\gamma_{QPC}$.

\bibitem{pdg} Particle Data Group,
Phys. Rev. D50, 1177/(1994). The data points in Fig. 2  are those in
that energy range in the graph
on p.1334. We thank Michael
Barnett and Tom Trippe for these selected data.
The caption to this graph reads in part:``Systematic normalization
errors are
not included; they range from $\sim$5-20\%, depending on experiment.  We
caution that especially the older experiments tend to have large normalization
uncertainties." Presumably BEPC will in future remeasure these cross
sections.
\bibitem{besson} D. Z. Besson et al., Phys. Rev. Lett. 381 (1985). The constant
non-\bb background in R has been subtracted from these published data using the
mean value 4.56 of the data points below beauty threshold, and the variance
of these points has been added in quadrature with the published error to
obtain the error bars in our graph.

\end{references}
\end{document}